\documentclass[superscriptaddress,preprintnumbers,amsmath,amssymb,twocolumn,floats]{revtex4-1}
\usepackage{bbm}
\usepackage{mathrsfs}
\usepackage{amsmath}
\usepackage{amsfonts}
\usepackage[colorlinks=true,citecolor=blue,anchorcolor=blue]{hyperref}
\usepackage{graphicx,epstopdf}
\usepackage{float}
\usepackage{subfigure}
\usepackage{epsfig}
\usepackage{dcolumn}
\usepackage{bm}
\usepackage{color}
\usepackage{natbib}
\usepackage{amssymb}
\usepackage{xcolor}
\usepackage{braket}
\usepackage{amsmath}
\numberwithin{equation}{section}
\begin{document}
\title{Absence of Altermagnetic Spin Splitting Character in Rutile Oxide RuO$_2$}

\author{Jiayu Liu}
\thanks{Equal contributions}
\affiliation{Shanghai Institute of Microsystem and Information Technology, Chinese Academy of Sciences, Shanghai 200050, China}
\affiliation{Shanghai Synchrotron Radiation Facility, Shanghai Advanced Research Institute, Chinese Academy of Sciences, Shanghai 201210, China}
\affiliation{University of Chinese Academy of Sciences, Beijing 100049, China}

\author{Jie Zhan}
\thanks{Equal contributions}
\affiliation{School of Materials Science and Engineering, University of Science and Technology of China, Shenyang 110016, China}
\affiliation{Shenyang National Laboratory for Materials Science, Institute of Metal Research, Chinese Academy of Sciences, Shenyang 110016, China}

\author{Tongrui Li}
\thanks{Equal contributions}
\affiliation{National Synchrotron Radiation Laboratory and School of Nuclear Science and Technology, University of Science and Technology of China, Hefei 230026, China}

\author{Jishan Liu}
\thanks{Equal contributions}
\email{liujs@sari.ac.cn}
\affiliation{Shanghai Synchrotron Radiation Facility, Shanghai Advanced Research Institute, Chinese Academy of Sciences, Shanghai 201210, China}
\affiliation{National Key Laboratory of Materials for Integrated Circuits, Shanghai Institute of Microsystem and Information Technology, Chinese Academy of Sciences, Shanghai 200050, China}

\author{Shufan Cheng}
\affiliation{National Laboratory of Solid State Microstructures and Department of Physics, Nanjing University, Nanjing 210093, China}

\author{Yuming Shi}
\affiliation{Shanghai Institute of Microsystem and Information Technology, Chinese Academy of Sciences, Shanghai 200050, China}
\affiliation{University of Chinese Academy of Sciences, Beijing 100049, China}

\author{Liwei Deng}
\affiliation{Shanghai Institute of Microsystem and Information Technology, Chinese Academy of Sciences, Shanghai 200050, China}
\affiliation{University of Chinese Academy of Sciences, Beijing 100049, China}

\author{Meng Zhang}
\affiliation{School of Physics, Zhejiang University, Hangzhou 310027, China}

\author{Chihao Li}
\affiliation{Laboratory of Advanced Materials, State Key Laboratory of Surface Physics, and Department of Physics, Fudan University, Shanghai 200438, China}

\author{Jianyang Ding}
\affiliation{Shanghai Institute of Microsystem and Information Technology, Chinese Academy of Sciences, Shanghai 200050, China}
\affiliation{University of Chinese Academy of Sciences, Beijing 100049, China}

\author{Qi Jiang}
\affiliation{Center for Transformative Science, ShanghaiTech University, Shanghai 201210, China}

\author{Mao Ye}
\affiliation{Shanghai Synchrotron Radiation Facility, Shanghai Advanced Research Institute, Chinese Academy of Sciences, Shanghai 201210, China}
\affiliation{National Key Laboratory of Materials for Integrated Circuits, Shanghai Institute of Microsystem and Information Technology, Chinese Academy of Sciences, Shanghai 200050, China}

\author{Zhengtai Liu}
\affiliation{Shanghai Synchrotron Radiation Facility, Shanghai Advanced Research Institute, Chinese Academy of Sciences, Shanghai 201210, China}
\affiliation{National Key Laboratory of Materials for Integrated Circuits, Shanghai Institute of Microsystem and Information Technology, Chinese Academy of Sciences, Shanghai 200050, China}

\author{Zhicheng Jiang}
\affiliation{National Synchrotron Radiation Laboratory and School of Nuclear Science and Technology, University of Science and Technology of China, Hefei 230026, China}

\author{Siyu Wang}
\affiliation{National Synchrotron Radiation Laboratory and School of Nuclear Science and Technology, University of Science and Technology of China, Hefei 230026, China}

\author{Qian Li}
\affiliation{National Synchrotron Radiation Laboratory and School of Nuclear Science and Technology, University of Science and Technology of China, Hefei 230026, China}

\author{Yanwu Xie}
\affiliation{School of Physics, Zhejiang University, Hangzhou 310027, China}

\author{Yilin Wang}
\affiliation{School of Emerging Technology, University of Science and Technology of China, Hefei 230026, China}
\affiliation{Hefei National Laboratory, University of Science and Technology of China, Hefei 230088, China}

\author{Shan Qiao}
\email{qiaoshan@mail.sim.ac.cn}
\affiliation{National Key Laboratory of Materials for Integrated Circuits, Shanghai Institute of Microsystem and Information Technology, Chinese Academy of Sciences, Shanghai 200050, China}

\author{Jinsheng Wen}
\email{jwen@nju.edu.cn}
\affiliation{National Laboratory of Solid State Microstructures and Department of Physics, Nanjing University, Nanjing 210093, China}
\affiliation{Collaborative Innovation Center of Advanced Microstructures, Nanjing University, Nanjing 210093, China}

\author{Yan Sun}
\email{sunyan@imr.ac.cn}
\affiliation{School of Materials Science and Engineering, University of Science and Technology of China, Shenyang 110016, China}
\affiliation{Shenyang National Laboratory for Materials Science, Institute of Metal Research, Chinese Academy of Sciences, Shenyang 110016, China}

\author{Dawei Shen}
\email{dwshen@ustc.edu.cn}
\affiliation{National Synchrotron Radiation Laboratory and School of Nuclear Science and Technology, University of Science and Technology of China, Hefei 230026, China}

\begin{abstract}

Rutile RuO$_2$ has been posited as a potential $d$-wave altermagnetism candidate, with a predicted significant spin splitting up to 1.4 eV. Despite accumulating theoretical predictions and transport measurements, direct spectroscopic observation of spin splitting has remained elusive. Here, we employ spin- and angle-resolved photoemission spectroscopy to investigate the band structures and spin polarization of thin-film and single-crystal RuO$_2$. Contrary to expectations of altermagnetism, our analysis indicates that RuO$_2$ ’s electronic structure aligns with those predicted under nonmagnetic conditions, exhibiting no evidence of the hypothesized spin splitting. Additionally, we observe significant in-plane spin polarization of the low-lying bulk bands, which is antisymmetric about the high-symmetry plane and contrary to the $d$-wave spin texture due to time-reversal symmetry breaking in altermagnetism. These findings definitively challenge the altermagnetic order previously proposed for rutile RuO$_2$, prompting a reevaluation of its magnetic properties.

\end{abstract}
\maketitle
\clearpage
Recently, altermagnetism (AM) has been proposed as a third fundamental magnetic phase beyond traditional ferromagnetism (FM) and antiferromagnetism (AFM) in crystals with collinear magnetic order~\cite{vsmejka2022beyond,savitsky2024researchers,vsme2022emerging}. In this magnetic phase, sublattices with opposite spins are connected by real-space rotation rather than translation or inversion transformation. This results in an unconventional AFM-like magnetic order with zero net magnetization, while exhibiting FM-like lifted Kramers spin degeneracy in reciprocal space [Fig. 1(a)]. This dichotomy allows altermagnets to combine the benefits of both ferromagnets and antiferromagnets, which were previously thought to be incompatible, offering advantages unparalleled by either conventional magnetic class. The unique properties of altermagnets could lead to the development of advanced materials with tailored magnetic and electronic properties, potentially revolutionizing magnetic storage and quantum computing~\cite{vsmejka2022beyond,bhowal2024ferroically,khalili2024prospects}. 

\begin{figure*}[tbp]
\includegraphics[]{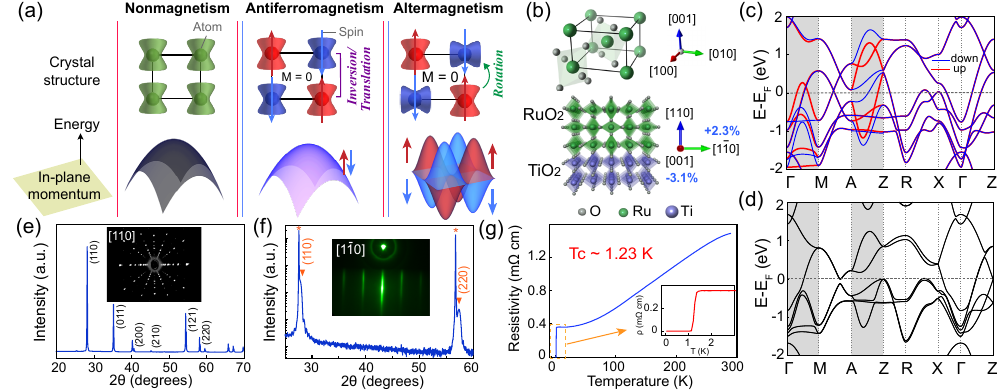}
\caption
{ 
The crystal structure, band structure, and characterizations of RuO$_2$. (a) The categorization of the magnetism of $\mathrm{RuO}_2$: nonmagnetism, antiferromagnetism, and altermagnetism. The crystal structures in real space and corresponding band structures in reciprocal space. (b) Crystal structure of $\mathrm{RuO}_2$, lower panel: Schematic diagram of the growth of thin films $\mathrm{RuO}_2$ on $\mathrm{TiO}_2$ (110) substrates. (c) Energy dispersion along the high-symmetry directions for the altermagnetic phase with a static $U$ = 1.3 eV correction on the $\mathrm{Ru}$ sites. (d) Energy dispersion for nonmagnetism of $\mathrm{RuO}_2$ with the consideration of spin-orbital coupling (SOC). (e) Powder XRD of single crystal $\mathrm{RuO}_2$. Inset displays the Laue pattern along the [110] direction. (f) $\theta$-$2\theta$ scan of $\mathrm{RuO}_2$ thin film grown on the (110)-oriented $\mathrm{TiO}_2$ substrate. Bragg peaks from the $\mathrm{TiO}_2$ substrates are marked with asterisks. The inset displays the RHEED pattern of the thin film. (g) Resistivity versus temperature curve of thin films.}
\label{NM state1}
\end{figure*}

Many theoretical and experimental studies have been conducted to predict and verify altermagnetic candidates in practical compounds~\cite{vsmejka2022beyond,bai2024altermagnetism}. Among them, rutile oxide RuO$_2$ is of particular interest and has been considered one of the workhorse materials of this emerging research~\cite{smolyanyuk2024fragility,sattigeri2023altermagnetic,ptok2023ruthenium,vsmejkal2022giant}. This interest is mainly rooted in predictions that rutile RuO$_2$ would exhibit the largest altermagnetic spin splitting, up to 1.4 eV, which could significantly enhance its potential for various spintronic applications by providing a stronger and clearer signal. Early resonant x-ray scattering~\cite{zhu2019anomalous} and neutron diffraction studies~\cite{berlijn2017itinerant} identified the AFM magnetic order in RuO$_2$. Subsequent transport measurements on RuO$_2$ revealed unconventional anomalous Hall and spin-polarized currents, which could be well rationalized within the altermagnetic framework~\cite{feng2022anomalous,vsmejkal2020crystal,gonzalez2021efficient,bai2023efficient,karube2022observation}. However, the latest independent muon-spin relaxation and rotation ($\mu$SR) studies, which are extremely sensitive probes to local magnetic moments, indicated an extremely small magnetically ordered moment of at most 1.4 $\times$ 10$^{-4}$$\mu_{\mathrm{B}}$/Ru for RuO$_2$ single crystals and 7.4 $\times$ 10$^{-4}$ $\mu_{\mathrm{B}}$/Ru for thin films, respectively~\cite{hiraishi2024nonmagnetic,kessler2024absence}. These significantly challenged previously assumed altermagnetic ground state in rutile RuO$_2$~\cite{smolyanyuk2024fragility,wenzel2024fermi}. 

\begin{figure*}[tbp]
\includegraphics[]{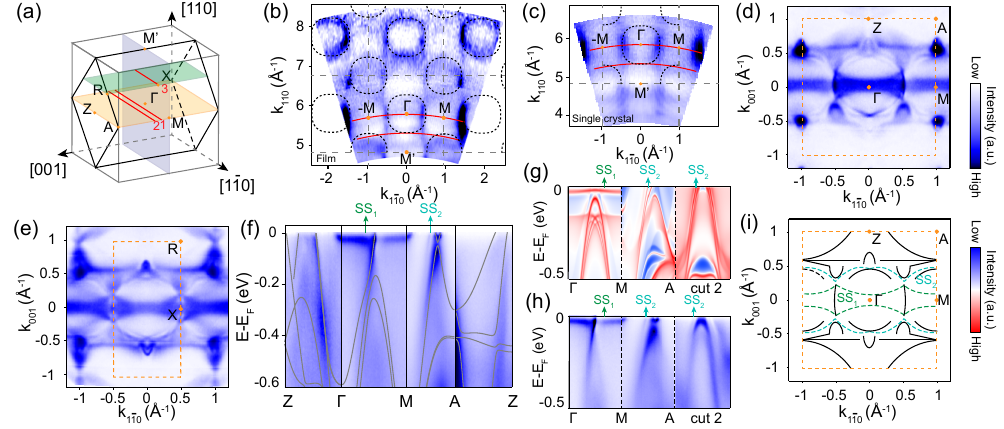}
\caption
{ 
Fermi surfaces and dispersions of RuO$_2$.
(a) 3D BZ of $\mathrm{RuO}_2$, with yellow and green planes showing high symmetry planes $\Gamma$--$M$--$A$--$Z$ and $R$--$X$. 
(b),(c) Photon emission intensity maps of thin films and single crystals along the out-of-plane direction, compiled over photon energy ranges from 82--300 eV and 50--160 eV, respectively. Gray dotted squares represent the BZ, and black dashed lines mark the Fermi surface for nonmagnetic calculations.
(d),(e) Fermi surface maps of the high symmetry planes $\Gamma$--$M$--$A$--$Z$ and $R$--$X$, measured at photon energies of 124 and 103 eV, respectively.
(f) Energy dispersions along high symmetry directions. Gray solid lines represent nonmagnetic calculated bulk bands. SS$_1$ and SS$_2$ are surface states.
(g),(h) Calculated surface band dispersions and the corresponding photoemission plots, respectively.
(i) Schematic of the Fermi surface, including bulk and surface states.
}
\label{NM state2}
\end{figure*}

Electronic structure probing techniques like angle-resolved photoemission spectroscopy (ARPES) and spin-resolved ARPES (SARPES) are capable of detecting momentum-dependent band splitting and spin polarization induced by AM~\cite{zhan2022angle,lv2019angle,schonhense2015space,okuda2017recent}. These techniques are thus considered the gold standard for identifying AM order in compounds. Recently, ARPES and SARPES measurements have been employed to verify AM ground states in several candidates, including MnTe, MnTe$_2$, and CrSb~\cite{ding2024large,yang2024three,reimers2024direct,krempask2024altermagnetic,zhu2024observation,lee2024broken}. Given the prominence and conceptual
significance of RuO$_2$ as a prototypical altermagnet, it is crucial to experimentally verify its altermagnetic nature from the perspective of electronic structure.

In this work, we performed ARPES and SARPES to examine the electronic structures of both thin-film and single-crystal rutile RuO$_2$. We show that the Fermi surface features and band structures along the high symmetry directions are well traced by density functional theory (DFT) calculations without magnetism. The band splitting expected from altermagnetism is not observed as well. Besides, we reveal significant in-plane spin polarization of the low-lying bulk bands, which is antisymmetric about the high-symmetry plane and clearly inconsistent with the expectation of altermagnetism. Our direct spectral evidence clearly shows that RuO$_2$ is highly unlikely
to be an altermagnet.

Figure 1(b) presents the crystal structure of rutile RuO$_2$ with the space group $\mathrm{P} 4_2 / \mathrm{mnm}$, where ruthenium(Ru) atoms are located at the centers of oxygen octahedra~\cite{occhialini2021local}. Neighboring oxygen octahedra are connected through rotation rather than translation, which would result in a typical $d$-wave altermagnetic ground state if the spins at Ru sites were aligned antiferromagnetically. 
Our first-principles calculations reveal spin splitting along the $\Gamma$---$M$ and $A$---$Z$ directions [Fig. 1(c)], consistent with previous predictions of the AM state~\cite{li2024observation,ah2019antiferromagnetism}. These bands are distinct from the nonmagnetic case shown in Fig. 1(d), where all bands are degenerate. Thus, directly investigating the low-lying band structure and corresponding spin polarization should be an effective method for identifying the magnetic ground state of rutile RuO$_2$. Given that previous $\mu$SR studies have thoroughly explored the magnetically ordered moments of both RuO$_2$ single crystals and thin films, we conducted ARPES and SARPES measurements on both types of samples to achieve comprehensive results. For detailed descriptions of sample syntheses, (S)ARPES measurements and first-principles calculations, please refer to the methods section in the Supplementary Material, note 1~\cite{SI_Here}. Note that the substrate-induced strain on the thin films does not alter the magnetic ground states according to our calculations (see details in Supplementary Material, note 2). Figure 1(e) shows the powder x-ray diffraction (XRD) and Laue patterns of RuO$_2$ single crystals. Figure 1(f) presents the high resolusion XRD and RHEED patterns of the epitaxial thin films. These demonstrate the high quality and atomically flat surface of the samples. Moreover, the superconducting transition temperature reaches up to 1.23 K [Fig. 1(g)], comparable to the highest recorded among all reported values~\cite{occhialini2022strain,ruf2021strain,uchida2020superconductivity}, further demonstrating the superior quality of our thin films for $in$--$situ$ ARPES and SARPES measurements.

A prerequisite for identifying possible altermagnetic
spin splitting in the low-energy electronic structure of
RuO$_2$ is the elimination of interference from complex
surface states~\cite{li2024observation}. To achieve this, we conducted detailed photon-energy-dependent ARPES measurements to distinguish surface and bulk states near the Fermi level ($E_F$). Figures 2(b) and 2(c) present photoemission intensity maps in the $k_{110}$-$k_{1\Bar{1}0}$ plane [the gray plane in the three-dimensional Brillouin zone (BZ) in Fig. 2(a)] taken at $E_F$ for thin films and single crystals, respectively. Both maps clearly show periodic alternation of large and slightly smaller circular Fermi pockets, which can be well reproduced by our nonmagnetic bulk band calculations (black dashed rings). Additionally, both maps exhibit nearly dispersionless chainlike features along the tangents of these pockets, which were not captured by our calculations, indicating their surface band nature.

\begin{figure*}[tbp]
\includegraphics[]{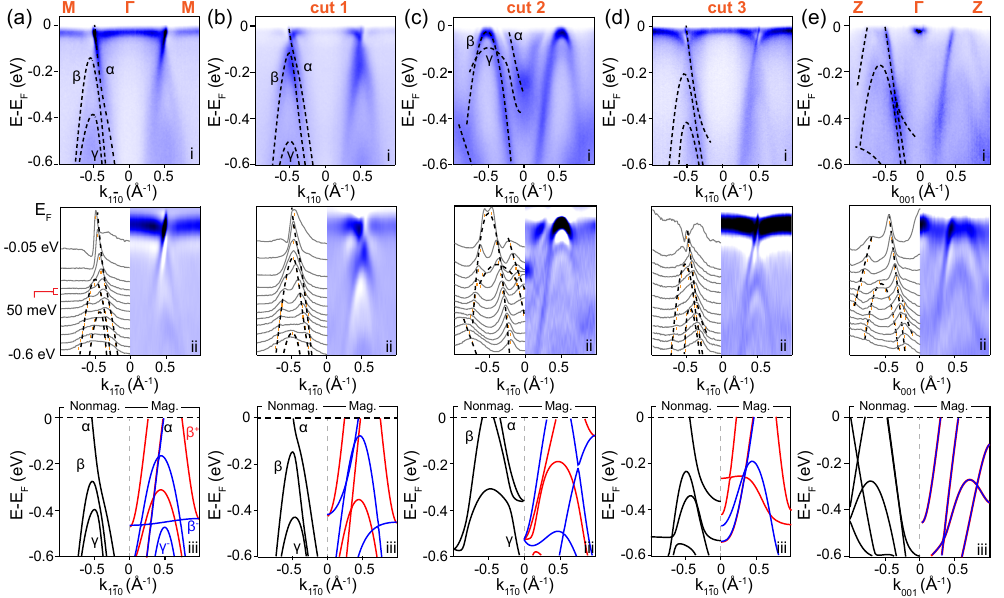}
\caption
{Comparisons of calculated energy dispersions in nonmagnetic and altermagnetic states.
(a)-(e) The photoemission plots at different momentum (i), momentum distribution curves, and the second derivative plots (ii), comparisons of nonmagnetic and magnetic calculations (iii).  Red and blue lines represent spin-up and spin-down states, respectively. Cut 3 was acquired at a photon energy of 103 eV, and the other four cuts at 124 eV.}
\label{NM state3}
\end{figure*}

\begin{figure*}[tbp]
\includegraphics[]{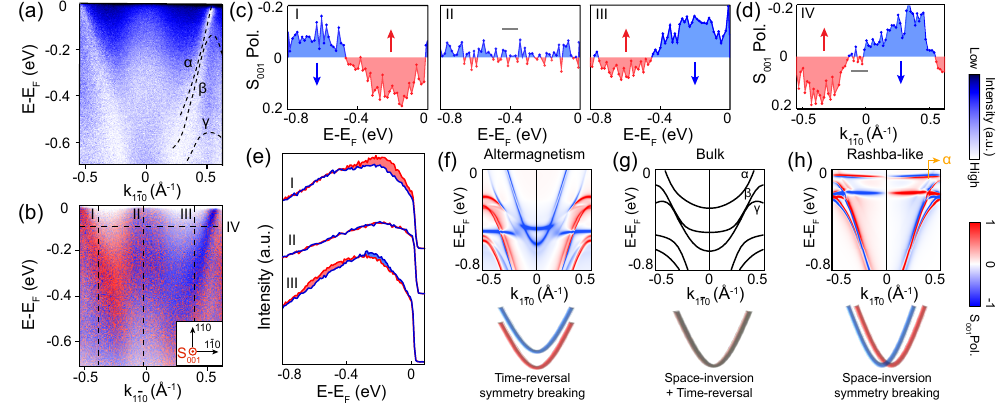}
\caption
{ Spin polarization in RuO$_2$.
(a) Spin-integrated band dispersion. (b) Spin polarization of the bulk band dispersion. 
(c) Spin polarization versus binding energy at different momenta indicated by the black dashed line in (b).
(d) Spin polarization versus momentum at the binding energy of around 0.1 eV. (e) Spin-resolved energy distributed curves at the three selected momenta. (f)--(h) Surface spin projected band structure of altermagnetic state (f), bulk band structure of nonmagnetic state (g), surface spin projected band structure of nonmagnetic state (h) and corresponding schematic diagrams of system’s symmetry.
}
\label{NM state4}
\end{figure*}

We investigated detailed the low-lying band structure on both high-symmetry planes $\Gamma$---$M$---$A$---$Z$ and $R$---$X$ [light yellow and light green planes shown in Fig. 2(a), respectively] to directly compare them to calculations. The corresponding photoemission intensity maps and plots taken from the single-crystal $\mathrm{RuO}_2$ are shown in Figs. 2(d)--2(f). Despite the complexity of the intertwined surface and bulk band structures, we could unambiguously recognize surface states (labeled as SS$_1$ and SS$_2$) by comparing theoretically determined surface band dispersions [Fig. 2(g)] with the corresponding photoemission plots [Fig. 2(h)]. It is crucial to pinpoint all surface states as the extra surface bands can easily be misinterpreted as AM-induced spin splitting. We reached the similar conclusion using thin film data (Supplemental Material, note 3) and schematically summarized our findings in Fig. 2(i), which are consistent with previous reports~\cite{jovic2018dirac,jovic2021momentum,jovi2019dirac}. We found that the bulk band dispersions along high-symmetry directions shown in Fig. 2(f) are in remarkable agreement with the nonmagnetic calculations (light gray lines), implying that RuO$_2$ is more likely a paramagnetic metal.  

We next examined the alignment of experimental bulk bands with magnetic and nonmagnetic calculations along high-symmetry directions $\Gamma$---$M$ [Fig. 3(a)], $\Gamma$---$Z$ [Fig. 3(e)], and three other typical cuts across the BZ [Figs. 3(b)--3(d), cuts 1--3 illustrated in Fig. 2(a)]. According to the proposed AM state in RuO$_2$, all these low-lying bulk bands should exhibit lifted spin degeneracy except for those along $\Gamma$---$Z$, which are connected by the screw rotation operation $S_{4z}=\{C_{4z}|(1/2,1/2,1/2)\}$. However, along $\Gamma$---$M$, we could only identify three bulk bands near the $E_F$: the electronlike $\alpha$ band crossing $E_F$, and two holelike $\beta$ and $\gamma$ bands with band tops located at $\sim$ 0.2 eV and 0.4 eV below $E_F$, respectively. Using momentum distribution curves (MDCs) and second derivative plot [Fig. 3(a) ii], we precisely extracted dispersions of three bands and appended them onto the photoemission intensity plot [black dashed line in Fig. 3(a) i]. No band splitting predicted by the magnetic calculations
was found in our data. We note the energy resolution of our experiments and emphasize that no spin splitting higher than 10 meV was observed (Fig. S4 of Supplemental Material). Figures 3(b) and 3(c) illustrate the evolution of three bulk bands with momentum. We observed that the $\alpha$ and $\beta$ bands move upwards in binding energy from the cut along $\Gamma$---$M$ to cut 2, while the top of the $\gamma$ band initially falls and then rises. This complex evolution of all three bulk bands can be qualitatively captured by nonmagnetic calculations, and no signs of the spin splitting suggested by the AM state calculations. Figure 3(d) is the photoemission intensity plot along cut 3, showing all three bulk bands similar to those along $\Gamma$---$M$, which is in good agreement with nonmagnetic calculations rather than AM predictions [Fig. 3(d) iii]. Figure 3(e) presents a comparison of our ARPES results along $\Gamma$---$Z$ with DFT calculations. Because of screw rotation symmetry protection, even the AM state prediction does not show lifted Kramer spin degeneracy [Fig. 3(e) iii]. Note that the band in the nonmagnetic calculations has been shifted to match the band dispersion [Fig. S5 of Supplemental Material]. We observed only one band crossing $E_F$ within the momentum range of 0 to 0.5 \AA$^{-1}$, which conflicts with the AM prediction of two bands crossing $E_F$. We performed a similar comparison on thin film, as detailed in Fig. S6 of Supplemental Material. Overall, our ARPES data, whether taken from single crystals or thin films, are more consistent with nonmagnetic calculations. 

In addition to using band splitting as a criterion, incorporating spin-polarized band characterization could provide more compelling evidence for determining the presence of an AM state~\cite{vsmejka2022beyond,krempask2024altermagnetic}. Therefore, we conducted SARPES experiments using a helium lamp with photon energy of 21.2 eV~\cite{zha2023improvement}. Figure 4(a) illustrates the spin-integrated ARPES result of thin-film RuO$_2$ taken along a cut parallel to $\Gamma$---$M$ while located close to the $X$---$R$ plane. The electronlike band could be identified as the bulk $\mathrm{\alpha}$ band[Fig. 4(g)] and referring to Fig. 3(d). The $S_{001}$-resolved photoemission intensity plot is shown in Fig. 4(b), in which this bulk band exhibit pronounced opposite spin polarizations for electrons of opposite momentum. Figures 4(c) and 4(e) display the $S_{001}$ polarization versus binding energy and spin-resolved energy distribution curves (EDCs) taken at three different momenta, respectively. Additionally, the $S_{001}$ polarization as a function of momentum at around 0.1 eV below $E_F$ is presented in Fig. 4(d). All these data suggest that the bulk band is highly spin polarized, with an in-plane polarization vector $S_{001}$ antisymmetric about the $\Gamma$---$Z$---$M'$ high-symmetry plane. This finding sharply contrasts with the mainstream prediction of the AM state, where the $\mathrm{\alpha}$ band should be fully spin polarized but symmetric about the high-symmetry plane, as suggested in Fig. 4(f) and Fig. S7. We emphasize that our SARPES findings are robust, as similar results were obtained from multiple thin-film and single-crystal samples (Supplemental Material, note 8). 

The spin polarization structure observed in RuO$_2$ is really intriguing for both space-inversion and time-reversal symmetries are preserved, and thus there should be no spin splitting, as schematically shown in Fig. 4(g). 
We speculate that the Rashba-like spin splitting may be due to the breaking of the space-inversion symmetry as the penetration depth of photoelectrons excited is
only around 10\ \AA\ in real experiments. Figure 4(h) presents the calculated spin polarization of low-lying bands contributed mainly from states in the top several layers of RuO$_2$, simulated through the Green's function calculations. Figure S9 also shows slab calculations that yield Rashba-like states. This simulation exhibits evident space-inversion symmetry breaking and is qualitatively consistent with our SARPES findings.  Furthermore, the presence of oxygen vacancies would break both the inversion symmetry of the space group and the central symmetry of the Ru site point group, which would inherently gift the bulk band structure with spin polarization attributes (Fig. S10 in Supplemental Material, note 1). In short, the observed Rashba-like splitting ambitiously ruling out the possibility of magnetism.

In summary, we utilize ARPES and SARPES to directly examine the band structures and spin polarization of thin-film and single-crystal rutile RuO$_2$ samples. Our results show that electronic structure of RuO$_2$ matches well with nonmagnetic conditions. We did not observe the momentum-dependent spin splitting that was expected from the AM state. Furthermore, we detect significant in-plane spin polarization in the low-lying bulk bands, which is antisymmetric relative to the high-symmetry plane and incompatible with the $d$-wave spin texture expected from time-reversal symmetry breaking in AM. Our work would stimulate further experimental and theoretical research in AM, underscoring the need for reconsidering both material candidates and theoretical models, while the observed unusual spin polarization may have a great potential application in spintronics. In addition, the anomalous spin polarization in RuO$_2$ may be closely related to the catalytic performance as RuO$_2$ has been proven to be a useful potential catalyst for energy conversion and storage applications.

~\\
\noindent\textbf{Acknowledgements}

This work is supported by the National Key R\&D Program of China (No. 2023YFA1406304), National Science Foundation of China (No. U2032208, No. 12004405, No. 52271016, No. 52188101), Anhui Provincial Natural Science Foundation (No. 2408085J003). J.S.W. is supported by National Key Projects for Research and Development of China (No. 2021YFA1400400) and National Natural Science Foundation of China (No. 12225407 and No. 12074174). Y.L.W. is supported by the Innovation Program for Quantum Science and Technology (No. 2021ZD0302800) and the National Natural Science Foundation of China (No. 12174365). Y.W.X. is supported by the National Natural Science Foundation of China (No. 12325402). S.Q. is supported by the National Natural Science Foundation of China (Grants No.
11927807 and No. U2032207). Z.C.J. acknowledges the China National Postdoctoral Program for Innovative Talents (BX20240348). Part of this research used Beamline 03U of the Shanghai Synchrotron Radiation Facility, which is supported by ME$^2$ project under Contract No. 11227902 from National Natural Science Foundation of China. Part of the numerical calculations in this study were carried out on the ORISE Supercomputer(No. DFZX202319).

\bibliographystyle{naturemag}

\end{document}